%% file: Ouyed-CSQCDIV-v1.tex
\documentclass[12pt]{article}
\usepackage{epsfig}
\usepackage{amsmath}

\input econfmacros-csqcd4.tex
\def\Title#1{\begin{center} {\Large {\bf #1} } \end{center}}

\begin{document}


\Title{The Burn-UD code for the numerical simulations of the Hadronic-to-Quark-Matter phase transition}

\begin{raggedright}

{Amir Ouyed, Luis Welbanks, Nico Koning, Rachid Ouyed \\
Department of Physics and Astronomy, University of Calgary, Calgary, Alberta,  T2N 1N4,
Canada\\

 {\tt Email: ahouyedh@ucalgary.ca}}
\bigskip\bigskip
\end{raggedright}

\section{Introduction}

Quantum-Chromo-Dynamics (QCD) has become increasingly important to astrophysics during the last few decades. QCD is becoming especially ubiquitous in the study of compact objects, most specifically in the study of neutron stars (NSs). 
The hypothesis that quark matter (specifically, matter  made of up, down, and strange quarks) could be the most stable state of matter in the universe \cite{PhysRevD.4.1601,PhysRevD.30.272} has revealed the possibility for the existence of compact objects  made entirely of deconfined quark matter namely, quark stars.  The extremely dense cores of NSs could allow for the possibility of QCD phase transitions  \cite{cabibbo1975exponential} leading to the formation of these quark stars.

 If the phase transition is an exothermic process, the energy could manifest itself observationally as an energetic event that could have drastic consequences to high-energy astrophysics.  The exothermic release of energy by a QCD phase transition in a NS, resulting in a quark star, may lead to a highly explosive astrophysical phenomenon known as the Quark Nova (QN) \cite{ouyed2002quark}. Calculations have revealed that the energy released could be as high as $10^{53}$ ergs \cite{ouyed2002quark}. Such excesses of energy could account for the mechanism behind poorly understood phenomena such as super-luminous supernovae \cite{ouyed2013sn}. Furthermore, the energetic ejecta of such events could trigger nuclear reactions leading to heavy element nucleo-synthesis \cite{jaikumar2007nucleosynthesis}, p-process element synthesis \cite{ouyed2014unified}, and spallation processes \cite{ouyed2014spallation,ouyedprl2011}. Thus proving that the transition from a NS to a quark star is an explosive one will have transformative impact to astrophysics.

However,  making use of QCD concepts  in the context of compact objects faces challenges. For one, the thermodynamic regimes appropriate for QCD in astrophysics are non-perturbative. For example, equations of states (EoSs) of nuclear/quark matter  that include QCD physics in  studies of compact stars cannot be constructed from first principles. Instead phenomenological EoSs are used.  Although these equations have been relatively successful in modelling certain aspects of compact stars,  different EoSs models have important distinctions that would manifest in different dynamics  in astrophysics, in particular in the interior of NSs. 

How do we choose amongst all the possible phenomenological equations of state? Perhaps astrophysics might be of help. Astrophysical observations that correlate with specific EoS models could constraint the correct equations of state. This in turn, would have consequences not only in astrophysics but in particle physics in general, for it would shed light onto the true nature of the non-perturbative regime of QCD. 

Perhaps, the first step in finding out the true equation of state would be in simulating the micro-physics behind the phase transition of a NS into a quark star. Such a simulation would cover the parameter space related to disparate EoS and different initial conditions. In doing so, one could accomplish two things: (i) simulate astrophysical signatures that could help discriminate between the parameters, and ultimately the equations of state, and (ii) shed light  on whether a QCD phase transition inside a NS could reach an explosive regime or the phase transition simmers quietly and slowly. 

Here we describe the Burn-UD code which has been built and used by the Quark-Nova group to investigate the physics of this transition.
Burn-UD \cite{niebergal2010numerical} is a hydrodynamic code that simulates the micro-physics of the hadronic-quark matter phase transition. The work in general explores the effects of different EoSs states on the transition,  the importance of the initial
conditions (prior to the transition), and the hydrodynamics behind the combustion process.  Other points are organized as
follows: Section \ref{model} lays out the assumptions used in our model. Section 3 is a very brief introduction to the Burn-UD code/software
 which is presented as a unifying platform for the QCD community  exploring the phases of Quark Matter and astrophysicists working on Compact Stars. Some preliminary results of simulations run using Burn-UD are discussed and examined in  Section \ref{results}. Section 5 looks at future work and finally   we  presents our conclusions in Section 6.

\section{Model Assumptions}\label{model}

Before we describe Burn-UD, we describe
 our assumptions concerning the  the stability of {\it uds} matter, introduction of the {\it s}-quark seeding, the fluidity of quarks, the reaction rates of the weak interaction, and the EoSs used.  

\subsection{Strange Matter Hypothesis}

The strange matter hypothesis states that matter made  of up, down and strange quarks could  be the most stable
state of matter \cite{PhysRevD.30.272}.  If true, then it would be interesting to investigate the
 feasibility of the  phase transition from hadronic to {\it uds} matter in the universe.   If this transition
 is energetic and if it leads to the formation of stable {\it uds} stars, the implications to astrophysics are undeniable.
 Here, we assume that the strange matter hypothesis is true. This is dynamically important because  it implies that once a critical amount of strange quarks are seeded in the core of a NS, the whole system will try to lower its energy by converting to the more stable strange matter.

\subsection{Seeding of Strange Matter}

There are numerous theories of how a {\it s}-quark  seed could be introduced to the core of a NS. Seeding has to appear in such a way that a critical number of strange quarks are created instantaneously.
In our model, we assume an initial seed. The cause of the seeding itself is left as an open question. Numerous studies have pushed forward different seeding models, such as clustering of hyperons in the core, and neutrino sparking \cite{alcock1986strange}. Another model explored for seeding is the creation of strange quarks through DM self-annihilation \cite{perez2013,perez2010dark}. This  seeding mechanism 
is further discussed in Section \ref{dmseed}.

\subsection{Asymptotic Freedom and the Fluidity of Quarks}

Because our model instantiates the combustion process in the core of the NS, we already assume a critical density for asymptotic freedom, i.e. hadrons in the core have deconfined into a fluid made of quarks. We therefore model the combustion of a fluid consisting  (ud)
matter  into a fluid made of  (uds) matter. Moreover, the timescale of the strong force, which governs quark confinement, is much shorter than the weak interaction, which is responsible of turning ud matter into uds matter. Thus, it is reasonable to assume that hadrons at the vicinity of the flame have already dissolved into their constituent up and down quarks \cite{niebergal2010numerical}.

The conversion process of ud into uds matter is modelled as a combustion process; the ud matter acts as a fuel that is burnt into a uds ash. However, usually combustion studies usually deal with a pre-mixed fuel, i.e. the fuel is mixed with the oxidant. In such a pre-mixed combustion, the speed of the flame is dictated by the thermal conductivity, and the hydrodynamics (diffusion and advection) become unimportant compared to the conduction of heat across the medium. 

However, in the case of the combustion of ud matter into uds matter, the oxidant (strange quarks instantiated in the core of the NS) is not pre-mixed. Thus, combustion is only activated when the strange matter oxidant diffuses into the ud fuel. When the strange quarks diffuse into the ud fuel, the up, down and strange quark mix will attempt to achieve chemical equilibrium, converting up and down quarks into strange quarks, releasing heat in the process.

Because the dynamics of combustion are dominated by a hydrodynamic process, i.e. diffusion of the strange quark oxidant into ud fuel, and not by heat transfer, it is necessary to solve numerically the full system of hydrodynamic equations. We solve the 1-D hydrodynamical equations:

\begin{equation}
\frac{\partial U}{\partial t}= - \nabla F (U) + S(U).
\end{equation}

where $ \nabla F (U)$ are the advection components, and $S(U)$ the source components. We refer the interested reader  to \cite{niebergal2010numerical} for the details of the hydrodynamics of the ud to uds combustion. 

\subsection{Reactions}\label{rates}

Once a strange matter seed is introduced, the following reactions will try to achieve equilibrium. This creates a net surplus of strange quarks, resulting in the conversion of the ud fluid into uds matter:

\begin{align}
 d &\rightarrow u + e^- + \overline{\nu}_e 
\\ u + e^- &\rightarrow d + \nu_e
\\ s &\rightarrow u + e^{-} + \overline{\nu}_e
\\ u + e^- &\rightarrow s + \overline{\nu_e}
\\ u + d &\rightarrow u + s
\end{align}

We take the rates for these reactions from \cite{anand1997burning}. See also \cite{niebergal2010numerical}.

\subsection{Neutrinos}\label{neutrino}

The {\it uds} mixture  will attempt to reach equilibrium through the weak interaction reactions. As such, the reactions will release neutrinos which  act as an energy sink that lowers the temperature behind the burning interface (see reactions above). The sink is parameterized and  is proportional to the neutrino mean free path. Se also \cite{niebergal2010numerical}.

\subsection{Equations of State and Color Superconductivity}

 We assume that both the ud and uds matter near the burning interface are deconfined. In its early version, Burn-UD
  utilizes the MIT bag model \cite{johnson1975bag} for both sides of the interface. The EoS used is:

\begin{align} \label{miteos}
P&= \frac{13}{36}\pi^2 T^4 + \frac{T^2}{2}\sum_i \mu_i^2+ \frac{1}{4 \pi ^2}\sum_i \mu_i^4-B 
\\ h &= 4(P+B)
\end{align}

Where $P$ is pressure, $T$ is temperature, $\mu_i$ is the chemical potential for quark specie $i$, $h$ is enthalpy, and $B$ is the Bag constant. 

In addition, as the {\it uds} phase (the ashes in our simulation) cools we assume that it becomes superconducting and
enters a  Color-Flavor-Locked (CFL) phase \cite{alford1999color}. 
We model the CFL phase through the MIT Bag Model EoS with the following prescription \cite{paulucci2008color}.

\begin{align}
\\ \Delta_0 &= 100 \text{ MeV}  
\\ T_c &= 2^{1/3}\times 0.57 \Delta_0
\\ \Delta &=2^{-1/3} \Delta_0 \sqrt{1- (T/T_c)^2}
\\ P &=P_{\text{non-CFL}} + \frac{3 \Delta ^2 \mu^2}{\pi^2} \label{eoscfl}
\\ h &= 4(P+B) - 2\times \frac{3 \Delta ^2 \mu^2}{\pi^2} 
\end{align}

\noindent where $T, P, h$, are temperature, pressure, enthalpy. $T_c$ is the critical temperature, where $T<T_c$ is the temperature for phase transition to CFL.  $ \Delta$ is the CFL energy gap,   and $ \Delta_0$ is the energy gap for CFL at $T=0$.   $P_{\text{non-CFL}}$ is defined in equation \eqref{miteos}. 

Thus, the CFL state stiffens the EoS by adding a gap term $ \frac{3 \Delta ^2 \mu^2}{\pi^2} $, and therefore effectively lowers the energy density of the system. This in turn releases energy that energizes the burning front. 

Previous studies, e.g. \cite{paulucci2008color}, usually model CFL in equilibrium requiring $\mu_s=\mu_u=\mu_d$ for the massless quark case. However we model the moment of the phase transition as a non-equilibrated system where $\mu_s <\mu_u$. Thus we add the following condition that dictates when the non-CFL to CFL transition happens:
\begin{align}
 \mu_u-\mu_s &< \Delta \label{mucond} \ ,
\\ T &< T_c \label{tccond} \ .
\end{align}

Thus, in a non-equilibrium situation,  $ \frac{3 \Delta ^2 \mu^2}{\pi^2} $ cannot simply be absorbed into an effective bag constant, because the CFL transition only appears when the conditions in equations \eqref{mucond} and \eqref{tccond} are met. 

CFL is triggered as follows: first the instantiated strange quark seed interacts with the ud fuel, raising $\mu_s$  but lowering $\mu_u$ and $\mu_d$. After some time at a point near the interface, the chemical potentials will almost equal each other, and condition \eqref{mucond} is met. At this point the EoS stiffens because the component   $ \frac{3 \Delta ^2 \mu^2}{\pi^2} $  is added to the pressure equation, which creates a large pressure differential across the interface that in turn, triggers an advection wave travelling from the uds ash to the ud fuel.

\section{Burn-UD} \label{code}

Burn-UD is a hydrodynamic combustion code used to model the phase transition of hadronic to quark matter. The most recent version of the code is written in java (the original was written in Fortran 95, \cite{niebergal2010numerical}) and will be publicly available soon on all major platforms (windows, linux, Macintosh).

Beyond the micro-physics modelled by Burn-UD that we described in the preceding sections, what makes Burn-UD distinct is that it has a user-friendly graphical user interface (GUI). The user can modify in a transparent and user-friendly way many  of the initial conditions, including initial temperature, distribution of the strange seed and initial total density. Fig. 1 shows a screenshot of the Burn-UD GUI in its most recent version. Furthermore, Burn-UD's GUI allows the user to change the EoS and turn on and off specific hydrodynamical effects such as advection or diffusion. 

 Burn-UD also contains graphic modules that plot in real time the different variables of the hydrodynamic simulation. The user can select to store the data for specific time intervals in order to seamlessly create an animation of the results. It also gives the user the ability to store the data in ascii files. 

	\begin{figure}
 \centering
         \includegraphics[width=0.9\textwidth]{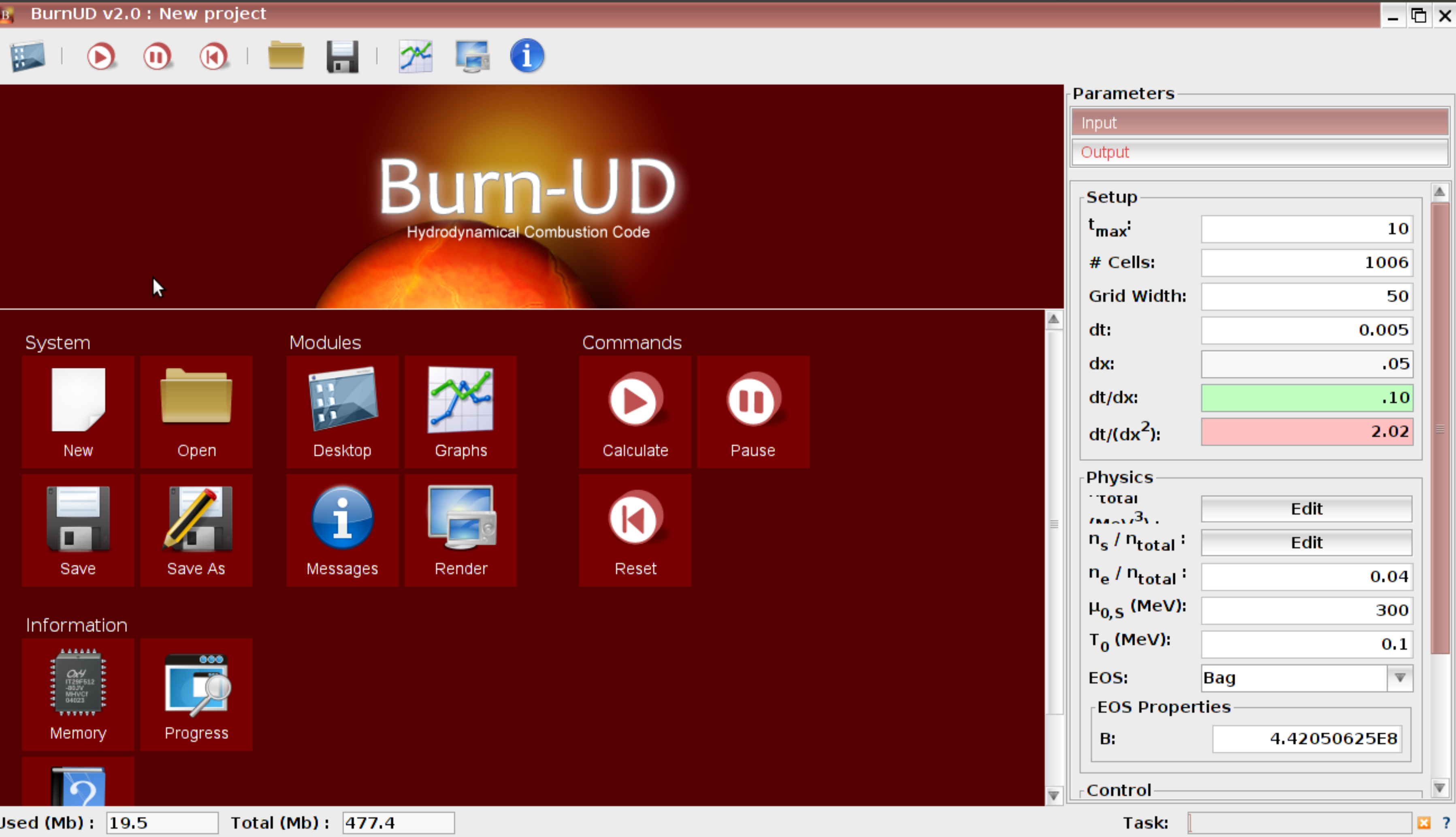}
    \includegraphics[width=0.9\textwidth]{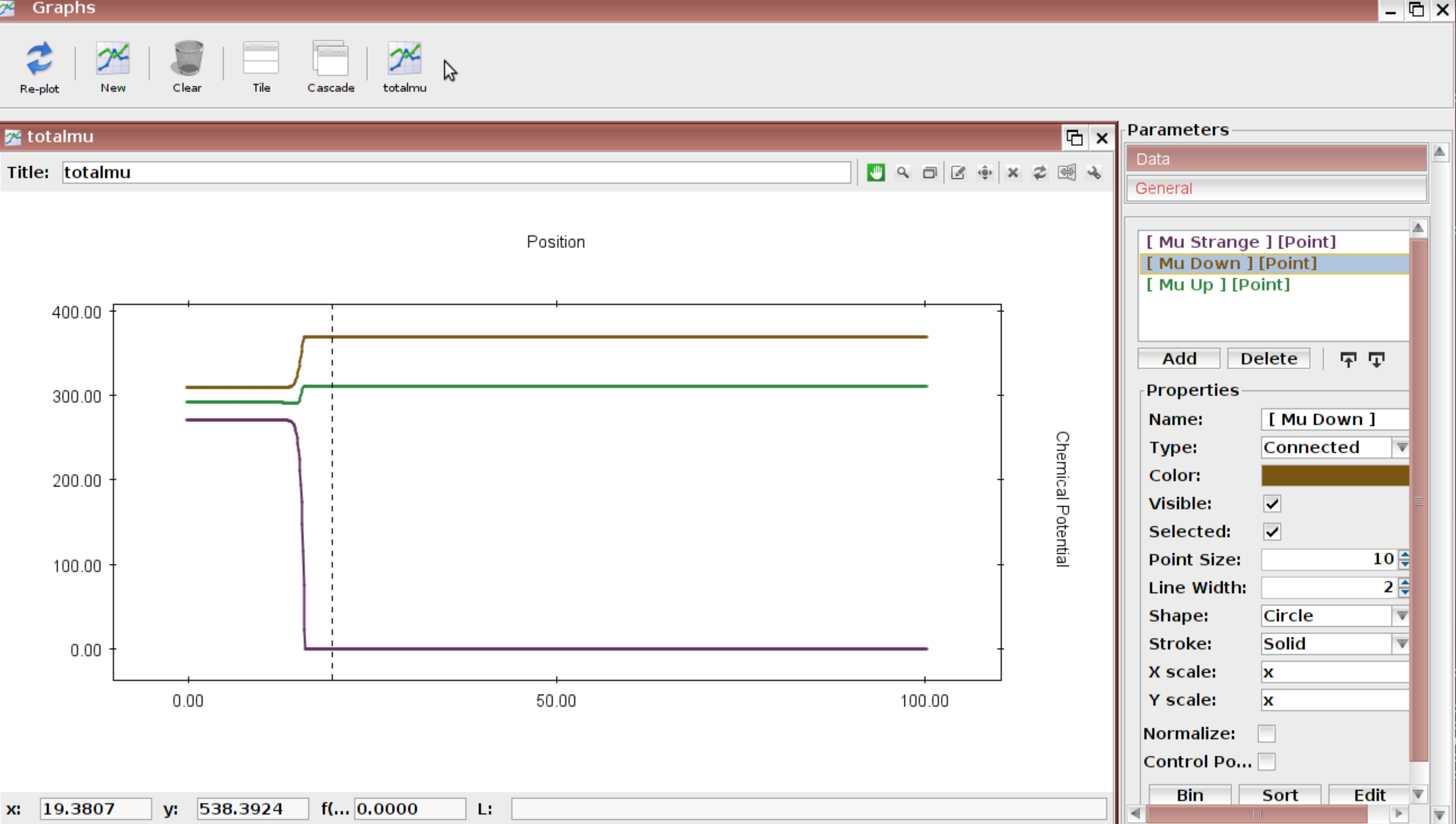} 
  \caption{ \label{code} Screenshots of the Burn-UD GUI. Upper window is the dashboard and the panel in the left is where the input parameters can be changed. Lower window is the graphing module where the chemical potentials of the strange, up, and down quarks are being plotted.}
\end{figure}

\section{Results}\label{results}

\subsection{Preliminary Results}

Early results using the first version of Burn-UD \cite{niebergal2010numerical} showed: (i) typical speeds for the burning front are 0.002c - 0.04c where c is the speed of light. (ii) Cooling due to escaped neutrinos slows down the burning front, and in some cases can halt it completely. 

These results hint towards an explosive regime. The details are explored in \cite{niebergal2010numerical} but can be summarized as follow: Neutrino cooling slows down the burning front unevenly across a multidimensional surface, which  wrinkles the interface. This wrinkling will be unstable, forming a deleptonization instability. This instability could either: (i) collapse the hadronic layers into the quark core and thus trigger a core collapse explosion or (ii) accelerate the burning front further and cause it to achieve supersonic speeds which would lead the system into a defragation to detonation regime, triggering an explosion.

\subsection{The CFL and Burning Speeds}

For the case of quark matter at a particular temperature and chemical potential, the CFL EoS has a lower energy density than the non-superconducting Bag equation of state used in our previous studies. Thus by simple energy conservation considerations, the transition of non-superconducting uds matter to CFL matter would release energy. This energy excess could also aid in leading the phase transition to an explosion.  Another important aspect of a CFL phase transition that could accelerate the burning is the release of a photon fireball \cite{ouyed2005fireballs}. Photons have momentum and thus, they will transfer momentum to the external hadronic layers that would in turn accelerate the burning.

\begin{figure}
 \centering
         \includegraphics[width=1.0\textwidth]{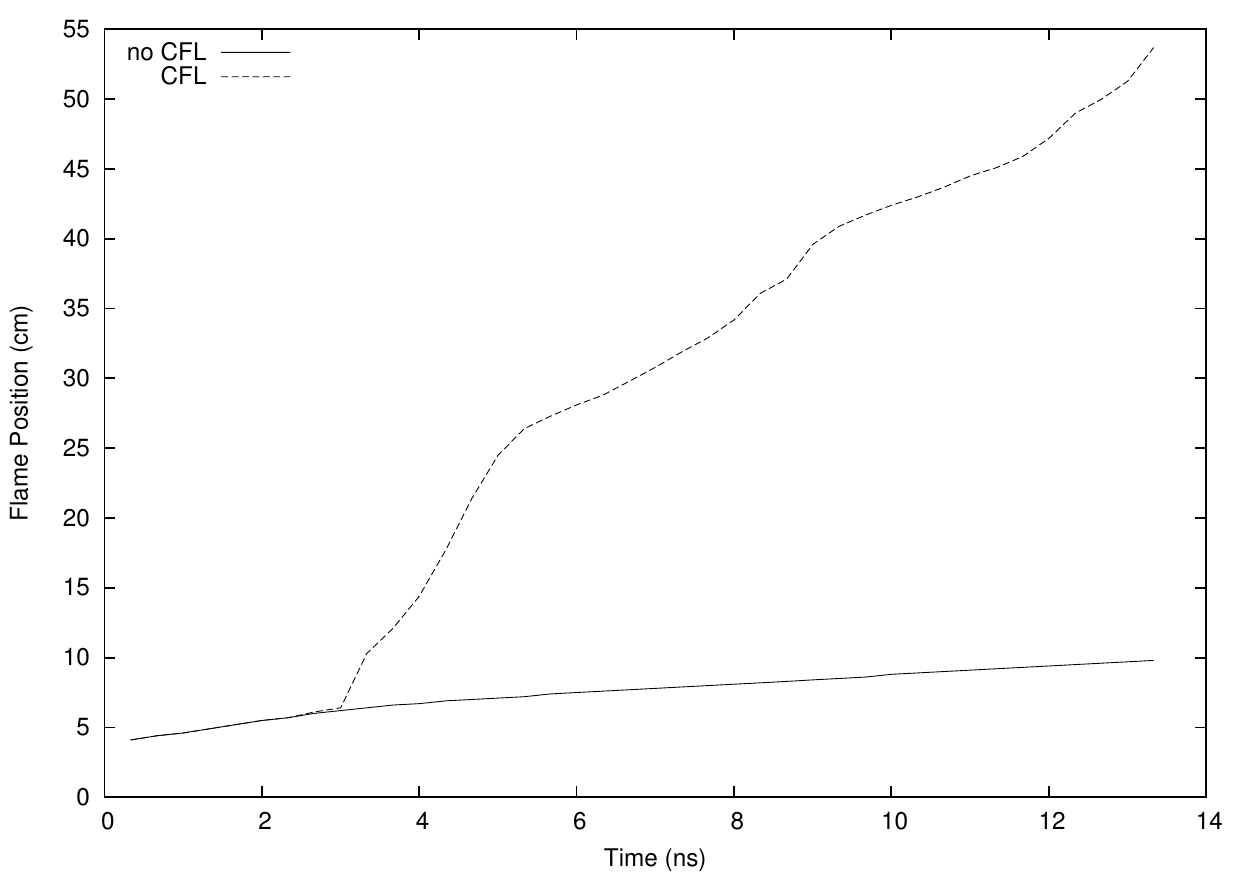}
  \caption{ \label{code} Position vs Time of the flame for a non-CFL (eq. \eqref{miteos}) EoS and a CFL (eq. \eqref{eoscfl}) EoS. Position is measured as the distance of the flame from the center of the NS.   The CFL flame is much faster than the non-CFL case. }
  	\label{fig:cfl}
\end{figure}

Several tests for the CFL case have been run, and the results seem promising. Figure \ref{fig:cfl} shows the 
position of the burning front versus time in nano-seconds. 
CFL stiffens the EoS which in turn creates an advection wave moving towards the unburnt ud fuel, which accelerates the burning front at even higher speeds.  The increase in speed at time $\sim 3$ nano-seconds in Figure \ref{fig:cfl} occurs when the
condition $(\mu_{\rm u} - \mu_{\rm s})< \Delta$  (see eq. (15)) is met. Furthermore, in a multidimensional situation,  the onset of CFL transition could further exacerbate the deleptonization wrinkling instability by further increasing the pressure gradient across the interface, by the simple fact that a CFL transition would increase the pressure in the cooler regions of the interface. Tests for the CFL case are ongoing to verify these claims.

\section{Future Considerations} \label{future}

\subsection{Dark Matter Annihilation and Strange Quark Seeding}\label{dmseed}

\begin{figure}
\centering
\includegraphics[width= 6 in]{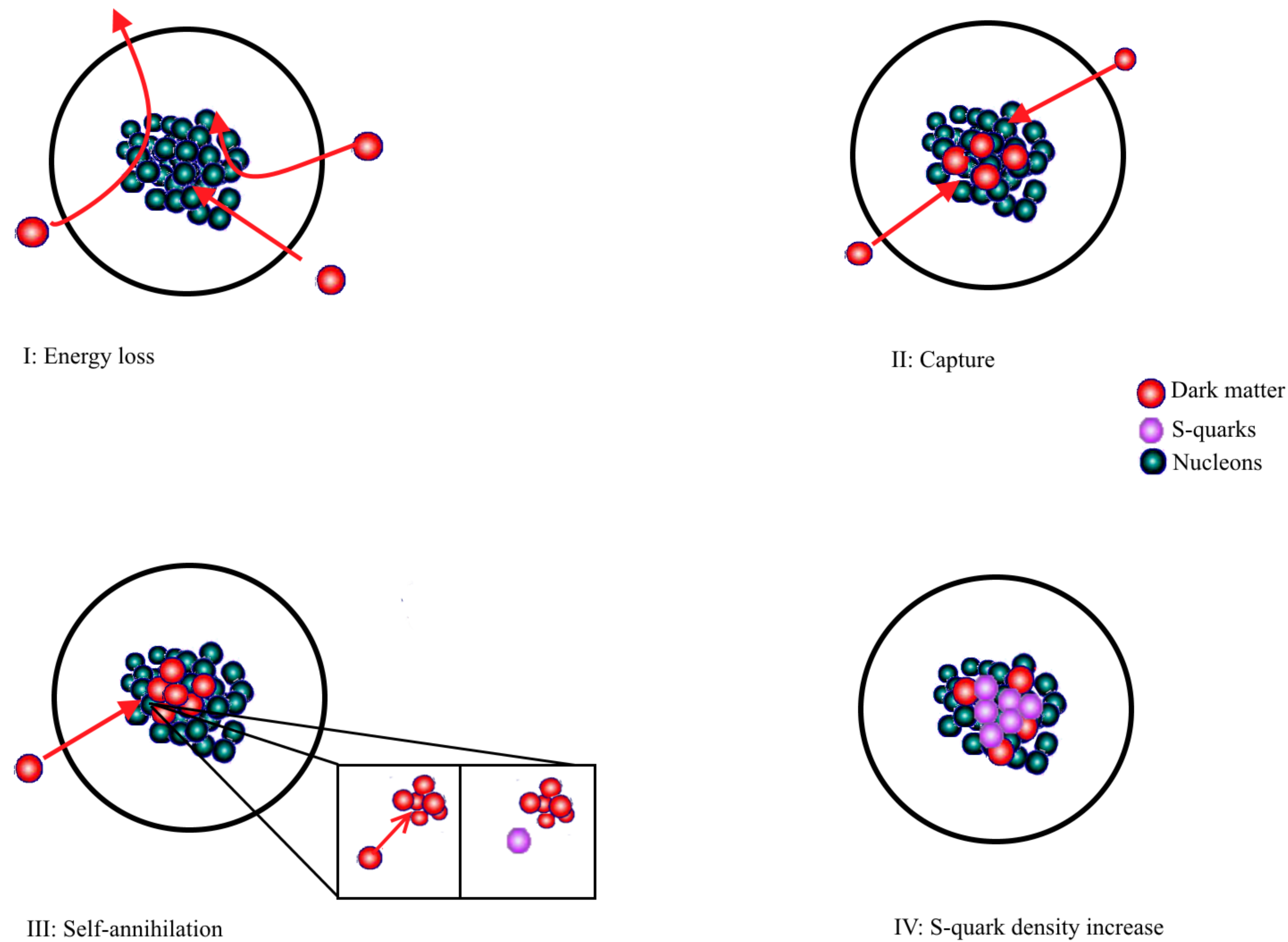}
\caption{Four main stages of DM capture by NSs. (I) Shows the DM energy loss by scattering with nucleons at high density inside the core of the NSs. (II) After losing energy, the DM particles are accreted in the core of the star and can't escape the gravitational pull of the star. (III) The density of DM has increased enough so that self-annihilation of DM happens in the core of the star. One of the possible products of this self-annihilation is the strange quark. Finally in (IV) the density of strange quarks increases, effectively seeding the hadronic-to-quark-matter phase transition process. }
\label{dmaccretion}
\end{figure}

A possible source of the strange quark seed may be dark matter (DM) annihilation \cite{perez2013,perez2010dark}. The distribution of DM  in galaxies allows for it to be accreted gravitationally onto a dense and massive astrophysical object such as a NS \cite{dmcapture}.  Figure \ref{dmaccretion} shows a schematic diagram representing the stages of DM accretion and self-annihilation to produce strange quarks.
One scenario explores the self-annihilation of  WIMPs (weakly interacting massive particles) inside NSs. They 
  inject the NS with energetic photons that in turn could deconfine hadrons into bubbles of ud quark plasma. Then non-leptonic weak processes could create strange quarks from these bubbles, giving rise to uds seeds \cite{perez2013,perez2010dark}. 
   Whether the amount of {\it s}-quark seeding is enough (i.e. reaches a critical mass) to trigger the hadronic-to-quark-matter phase
   transition remains to be confirmed making Burn-UD an ideal tool/software for such investigations.

Burn-UD is flexible enough to be used as a tool to study the strange quark seeding produced by DM annihilation. Burn-UD allows the user to input their own function for strange matter initial distribution.  Different seeding models might imply different functions for the initial distribution of the strange matter seed. The fact that Burn-UD allows the user to input their own   strange matter seed initial distribution makes Burn-UD useful for the study of many seeding models, including models for DM.

\subsection{Mass of the Strange Quark}

Up and down quarks have similar masses, yet the strange quark has a mass that is an order of magnitude higher. This difference between the up, down and strange mass should have consequences in the dynamics of the combustion.
At the moment, the EoS used for Burn-UD assumes the quarks are massless. The only dependency in Burn-UD to the quark mass lies in the rates mentioned in section \ref{results}, where a strange quark mass of $150 \text{ MeV}$  only affects the results very weakly as explored in \cite{niebergal2010numerical}.

Different masses for quarks would affect the binding energy necessary for the assembling of quarks. Therefore the creation of massive quarks would deplete the burning front of energy and slow down the burning speed.  It is important to include the quark masses in future studies to investigate how the results of the Burn-UD simulations are affected.

\subsection{More Sophisticated Equations of State}

Burn-UD models the flame micro-physics by imposing the MIT Bag EoS on both sides of the interface, i.e. for both the uds ash and the ud fuel. Thus the EoS at both sides of the interface shares the $B$ parameter as well. This means that at the region near the interface, where quarks have already deconfined,  the behaviour of the fluid becomes independent of the $B$ parameter. The speed of the burning front then becomes a function of the binding energy released during the creation and destruction of quarks. 

The MIT Bag Model is very popular because despite its simplicity, it manages to  successfully reproduce some of the properties of hadrons. It also models asymptotic freedom in an intuitive way making it useful for simulating the properties of bulk quark matter.
Yet such simplifications might miss some important dynamics.  One of them is chiral symmetry breaking.  A popular model that includes chirality is the Nambu-Jonas-Lasinio (NJL) model for QCD. One important difference in the NJL picture, is that strange quark matter is not the ground state of all matter \cite{buballa2005njl}. However, the NJL model, like  the MIT Bag Model, is phenomenological and as such its predictions are not completely conclusive either. In future releases of Burn-UD, the code will incorporate several EoS's, including NJL, to compare their effects on the simulation.

Finally, a more realistic simulation would include a mixed phase EoS as well. At the moment, deconfined quarks that act as a fluid are only considered. However, the dissolution of hadrons into quarks, if it is a first order phase transition, would change the pressure and therefore preshock the ud fuel to a certain degree, which would change the dynamics of the combustion simulation. 

\subsection{More Sophisticated Neutrino Transport}

As shown in \ref{neutrino}, Burn-UD currently models neutrino transport  as a parameterized energy sink that saps the matter near the interface of energy. Although this model addresses many of the larger and more qualitative aspects of neutrino transport, a more accurate model would actually implement the time dependent Boltzmann equations of neutrino transport. In the Boltzmann equations, neutrino flux is modelled as a density function that evolves across time and space. 

Having an accurate model of the neutrino transport is vital because the dynamics of these neutrinos are directly related to the deleptonization instability occurring at the burning interface. A sophisticated neutrino transport model will be one of the deciding ingredients in the discovery of whether the deleptonization instability leads to supersonic deflagration to detonation burning, or core-collapse Quark-Nova.

\section{Conclusion}\label{conclusion}

We hope that Burn-UD will evolve into a platform/software to be used and shared by  the QCD 
community exploring the phases of Quark Matter and astrophysicists working on Compact Stars. 
Burn-UD offers the possibility of studying the time-dependent phase transition of hadronic matter to quark matter. This software models hydrodynamically the combustion front (i.e. the flame's dynamics) which gives a physical window to diagnose whether the combustion process will simmer quietly and slowly, or transitions from deflagration to detonation. However, it is important to note that there exists solutions where the burning front is halted due to the deleptonization instability which make the system undergo a core-collapse QN explosion.  
Whether via a Deflagration-Detonation-Transition  or core-collapse, such an energetic phase transition would have many consequences in high-energy astrophysics and could aid in our understanding of many still enigmatic astrophysical transients. Furthermore, having a precise understanding of the phase transition dynamics for different EoSs could aid further in constraining the nature of the non-perturbative regimes of QCD in general.

\end{document}

%% file: econfmacros-csqcd4.tex
\def\beq{\begin{equation}}
\def\eeq#1{\label{#1}\end{equation}}
\def\eeqn{\end{equation}}

\def\beqa{\begin{eqnarray}}
\def\eeqa#1{\label{#1}\end{eqnarray}}
\def\eeqan{\end{eqnarray}}

\let\bar=\overbar

\def\Dslash{\not{\hbox{\kern-4pt $D$}}}
\def\dslash{\not{\hbox{\kern-2pt $\del$}}}

\def\msb{{\bar{\ssstyle M \kern -1pt S}}}

\usepackage{fancyhdr,graphicx}